\pdfoutput=1
\documentclass{article}
\usepackage{arxiv}
\usepackage{comment}
\usepackage{color}
\usepackage[utf8]{inputenc} 
\usepackage[T1]{fontenc}    
\usepackage{hyperref}       
\usepackage{url}            
\usepackage{booktabs}       
\usepackage{amsfonts}       
\usepackage{nicefrac}       
\usepackage{microtype}      
\usepackage{lipsum}
\usepackage{graphicx}
\usepackage[inline]{enumitem}
\newcommand\Mark[1]{\textsuperscript#1}
\usepackage{etoolbox}
\newcommand*{\affmark}[1][*]{\textsuperscript{#1}}

\title{
A Stance Data Set on Polarized Conversations on Twitter about the Efficacy of Hydroxychloroquine as a Treatment for COVID-19
}

\author{
  Ece Çiğdem Mutlu\affmark[1,]\thanks{All authors contributed equally to this study.}, Toktam Oghaz\affmark[2], Jasser Jasser\affmark[2], Ege Tütüncüler\affmark[1] \\
  \textbf{Amirarsalan Rajabi\affmark[2], Aida Tayebi\affmark[1], Ozlem Ozmen\affmark[1], Ivan Garibay\affmark[1,2,]\thanks{Corresponding author: Ivan Garibay, igaribay@ucf.edu}} \\
 \Mark{1}Department of Industrial Engineering\\
 \Mark{2}Department of Computer Science\\
  \texttt{\{ece.mutlu, jasser.jasser, ege.tutunculer, igaribay, ozlem\}@ucf.edu} \\
  \texttt{\{toktam\}@cs.ucf.edu, \{amirarsalan, aida.tayebi\}@knights.ucf.edu} \\
}

\begin{document}
\maketitle
\begin{abstract}
At the time of this study, the SARS-CoV-2 virus that caused the COVID-19 pandemic has spread significantly across the world. Considering the uncertainty about policies, health risks, financial difficulties, etc. the online media, specially the Twitter platform, is experiencing high volume of activity related to this pandemic. Among the hot topics, the polarized debates about unconfirmed medicines for the treatment and prevention of the disease have attracted significant attention from online media users. In this work, we present a stance data set, COVID-CQ, of user-generated content on Twitter in the context of COVID-19. We investigated more than 14 thousand tweets and manually annotated the opinions of the tweet initiators regarding the use of ``chloroquine'' and ``hydroxychloroquine'' for the treatment or prevention of COVID-19. To the best of our knowledge, COVID-CQ is the first data set of Twitter users' stances in the context of the COVID-19 pandemic, and the largest Twitter data set on users' stances towards a claim, in any domain. We have made this data set available to the research community via GitHub\footnote{\url{https://github.com/eceveco/COVID-CQ}}. We expect this data set to be useful for many research purposes, including stance detection, evolution and dynamics of opinions regarding this outbreak, and changes in opinions in response to the exogenous shocks such as policy decisions and events. 
\end{abstract}

\section{Introduction}
By August 2020, about 20 million confirmed infected cases with SARS-CoV-2 virus have been reported worldwide. Among the affected countries, the United States of America has reported the highest amount of infection, which is about 5 million infected individuals. The rapid spread of the virus and the uncertainty and risks associated with the COVID-19 pandemic has led institutions, policymakers, and individuals to seek drastic countermeasures against the spread of this disease, while imposing the least costs \cite{rajabi2020investigating}. For instance, many governments have taken severe responses to contain the virus, such as lock down the infected regions and shuttering their economies for weeks, closing their borders, and investing an unprecedented amount of funding on medical facilities and equipment.  In the absence of a vaccine, healthcare professionals resorted to alternative uses of existing drugs. Examples of such drugs are chloroquine (CQ) and hydroxychloroquine (HCQ), which are immunosuppressive and anti-parasite drugs that have been in use to treat malaria and lupus. Hydroxychloroquine is a less toxic metobolite of chloroquine and has been identified to have less side effects \cite{stokkermans2019chloroquine}. Both of these drugs are included in the treatment regimen of COVID-19 patients by physicians in Italy, France, and China, after some studies suggested that their use could be effective in inhibiting COVID-19 infections \cite{liu2020hydroxychloroquine, principi2020chloroquine}. 

While the majority of these studies came under heavy criticism due to the lack of scientific rigor, such as unusually small sample sizes and the absence of randomized trials, certain doctors, physicians and politicians quickly embraced the idea that hydroxychloroquine could be a “miracle cure”. On the other hand, some studies have cautioned against the use of chloroquine and hydroxychloroquine for the treatment of COVID-19 due to their hazardous side effects and their inefficacy \cite{sachdeva2020hydroxychloroquine,mahevas2020no}. In this information-scape of conflicting results and uncertainty, the unproven claim that Chlroquine and hydroxychloroquine are cure for COVID-19 quickly spread in online social networks and mainstream media outlets. In the White House press briefing dated March 19, 2020, hydroxychloroquine received endorsement from the US Administration, which accelerated spread of the rumors concerning the effectiveness of hydroxychloroquine against the disease. Thus, extensive polarization is observed on the effectiveness of these two drugs, specially, hydroxychloroquine, while this polarization is fueled by societal, political, and medical discussions about these drugs. 

Following these events and the political attention that hydroxychloroquine received, the claim that chloroquine and hydroxychloroquine may be effective tools against COVID-19 created tensions in social media, with many people posting about evidence for or against the effectiveness of both drugs, supporting or rejecting their use in a political tone, or just making neutral remarks about the ongoing chloroquine/hydroxychloroquine and COVID-19 situation. In this study, we aim to identify and analyze Twitter users’ stances about the rumor that chloroquine and hydroxychloroquine are cure for the novel coronavirus. For this work, we define rumor as an unofficial story or piece of news that might be true or invented, and quickly spreads from person to person, per Cambridge Dictionary. Our contributions are as follows:
\begin{itemize}
    \item We  present a manually annotated Twitter stance  data set for the unproven claim of ``chloroquine/hydroxychloroquine are cure for the novel coronavirus'', that we call COVID-CQ and it is accessible to the public from our GitHub repository\footnote{\url{https://github.com/eceveco/COVID-CQ}}. 
    \item To the best of our knowledge, COVID-CQ is the first stance dataset regarding COVID-19. This data can be used by the research community to analyze the dynamics of the opinions of social media users in a worldwide pandemic. 
    \item To  the  best  of  our knowledge,  COVID-CQ  is  currently  the  largest  human-labeled stance  dataset  on Twitter  conversations  with  more than  14 thousand stance labels towards a claim.
    \item The conducted annotation procedure includes joint annotation of tweets and shared URLs for disambiguation. This technique allowed us to produce accurate annotations for a challenging dataset that requires deep understanding of the content to identify the true stance of a tweet. 
    
\end{itemize}

\section{Related Work}
Detecting opposite opinions in a polarized conversation towards a target subject is a sub-domain of sentiment analysis, usually referred to as stance classification \cite{hacohen2017stance}. In stance classification, the subject under investigation is generally a person, organization, policy, or opinion \cite{sobhani2017stance}. The importance of this field of research relies on its applications, including the automatic extraction of attitudes and opinions towards events, and fake news or rumor identification \cite{aker2017simple}. Many studies on the literature of fact checking consider applying stance classification on potentially related documents to gain knowledge on varying sources of information, and predict the factuality of a claim according to the strength of aggregated stances \cite{baly2018integrating}. Generally, three stance classes have been considered in the literature for the task of stance classification: 
\begin{enumerate*}[label=\roman*)]
    \item positive (in favor), 
    \item negative (against), and
    \item neutral (none or neither);
\end{enumerate*}
 however, some studies have approached this classification by considering an extra class for the unrelated or irrelevant documents \cite{augenstein2016stance}. Despite the common categorization of stances as mentioned earlier, the idea of relative classification of stances using a ranking mechanism is proposed in \cite{zhang2018ranking}. The problem of stance classification was further evolved by the introduction of a complementary task to stance classification, which focuses on the stances towards given claims rather than entities, proposed in \cite{bar2017stance}. The first data set on stance classification towards claims was also proposed in \cite{bar2017stance}, which introduced a stance data set for 2,394 claims on Wikipedia articles. This data set is the most similar corpus to our data set. However, our focus is on the Twitter platform. 

Despite the existence of many stance data sets for polarized social media subjects, the available data sets are mostly focused on multiple target subjects, which results in having a small number of samples for each of the stance classes. For instance, the data set introduced in \cite{ferreira2016emergent} only contains labels for 4,455 tweets regarding the US presidential candidates for the 2016 election: ``Donald  Trump'',  ``Hillary  Clinton'',  ``Ted  Cruz'', and ``Bernie Sanders'' . Also, the SemEval-2016 Task 6 benchmark \cite{mohammad2016semeval} only contains 4,870 tweets with stance labels for 6 targets:  ``Atheism'',  ``Climate  Change  is  a  Real  Concern'', ``Feminist  Movement’',  ``Hillary  Clinton'', ``Legalization of Abortion'', and ``Donald Trump''. In addition to having a small size, SemEval data set was collected via hashtag queries and only specific tweets in which the hashtag appeared at the end of the tweet were considered. However, we did not exclude any tweets from our corpus. Our selection criteria is a simple keyword query of root. Another related data set to our work is the Stances in Replies and Quites (SRQ) data set proposed in \cite{villa2020stance}, which contains stance labels for approximately 5,200 replies and quotes to root tweets regarding varying events, including: ``Student Marches'', ``Iran Deal'', and ``Santa Fe Shooting''. The largest stance data set for Twitter is proposed in \cite{bar2017stance}, which contains stance labels for 51,284 tweets regarding 5 operations related to the health and entertainment industries. However, this data set does not investigate stances toward any particular claims.

Another related data set to our work is the TweetsCOV19 benchmark proposed in \cite{dimitrov2020tweetscov19}, which contains the available knowledge base of more than 8 million tweets, including the metadata of the tweets, besides the extracted entities, hashtags, user mentions, sentiments, and URLs. In contrast with our work, this data set does not contain user stances towards a claim or an entity. 

Our proposed corpus differs from the reviewed data sets on various aspects. First, we only focus on the stances of Twitter users towards a single claim of ``chloroquine and Hydroxychloroqine are cure for the novel coronavirus''. Second, focusing on a single claim, our data set can be used to investigate the dynamics of opinions over time towards the use of these drugs, in response to the exogenous shocks such as academic publications and events. Thus, our data set fills the gaps between the changes of opinions over time and stance classification, which is not possible to investigate using existing stance data sets. Third, considering the challenges regarding the true inference of the underlying stance in a short text, we approached the problem of stance annotation as a joint labeling of tweet and shared URLs if the tweet is not self-explanatory. 
For instance, the Twitter platform imposes a maximum length of 280 characters for each tweet, which leads to the ambiguity of short tweets and the challenges for the inference of the true stances. As a result of this work, we aim to introduce new challenges to the field of artificial intelligence, particularly, to the design of stance data intensive classification models, and to encourage the design of algorithms that consider the utilization of all sources of information with the goal of achieving accurate stance classification of textual content. 

\section{Stance Annotation}
Eisenberg and Finlayson define annotation as the "process of explicitly encoding information about a text that would otherwise remain implicit" \cite{eisenberg2019annotation}. For this study, annotation is the record of the Twitter audiences' stances about the running debate of chloroquine and hydroxychloroquine as treatments for coronavirus. Our purpose is to create a pure human-annotated data set rather than ML-based data set with or without supervision, as we believe that human-annotation renders further studies more robust and reliable. The annotation procedure was conducted by a team of containing 6 graduate and 3 undergraduate students in order to reach a consensus on the annotation guideline. In our annotation procedure, each student was asked to annotate the individual tweets as \textit{"Against","Favor" or "Neutral/None"} for the unproven claim of \textit{"chloroquine/hydroxychloroquine is cure for the novel coronavirus"}, relying on the well-known rumor listing website of fullfact.org \footnote{https://fullfact.org} as in Kwan et al.’s study \cite{kwon2017rumor}. We have done our analysis on online users’ conversations regarding topics related to the COVID-19 pandemic in the Twitter platform. The data was collected using the Twitter API and Hydrator \footnote{https://github.com/DocNow/hydrator} as suggested in \cite{chen2020tracking}. The detailed list of keywords being used for our data collection is given in the cited study. We considered tweets only related to the specific rumor and filtered tweets which include \textit{"hydroxychloroquine, chloroquine, and HCQ"} as keywords for our queries, published between 04/01/2020 and 04/30/2020. This data set includes 98,671 tweets generated by 75,685 unique Twitter users. Since stances of the retweets may be easily attained with assumptions, we focused only on the 14,374 unique tweets (tweets/mentions/replies) generated by 11,552 unique users (The most active user has 91 user-generated contents.) to decrease the work-load.

\begin{figure}[ht]
\centering
\includegraphics[width = .8\linewidth]{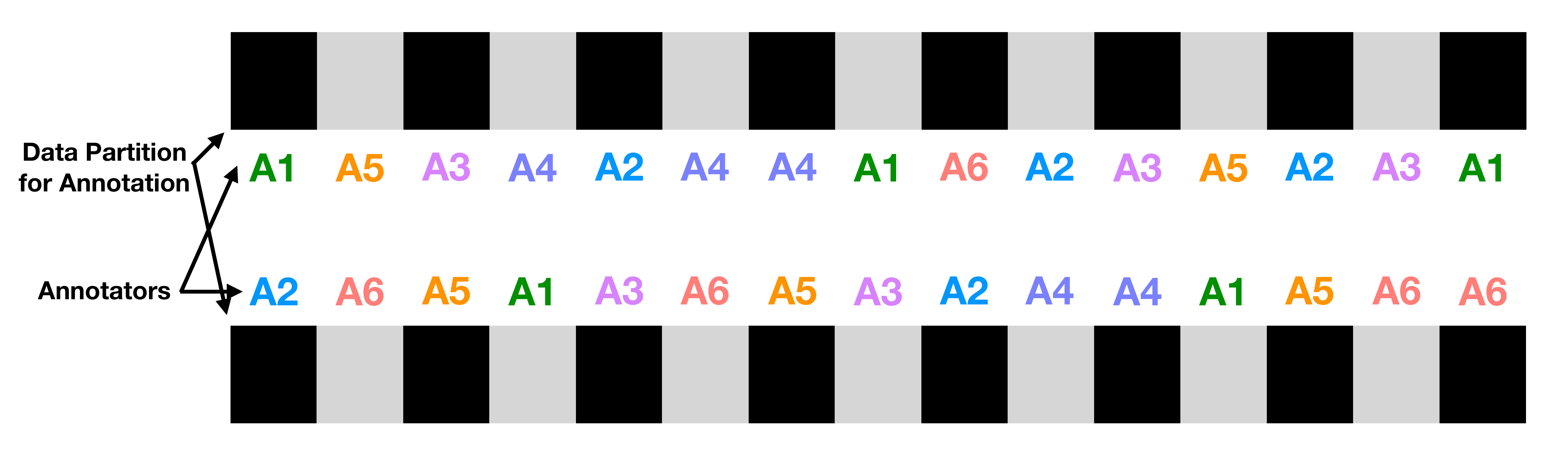}
\caption{A visualization of the distribution of our data set for stance annotation.}
\label{fig:annotators}
\end{figure}

Each of the investigated tweets is annotated by at least two different annotators. In the first round, the data was partitioned into 15 different clusters according to the time of the information creation and the tweet clusters were randomly assigned to the every possible combination of our annotator pairs. Thus, any possible biases due to annotator-pair match and time are prevented (Figure \ref{fig:annotators}). Among all the performed annotations, the inter-annotator agreement on this set was 87.37\%, which demonstrates that our annotators were effectively educated before conducting the procedure. In the second round, the remaining 12.63\% of the data was assigned to a third annotator who was not being asked to annotate that specific tweet for the first round. After labeling all tweets, the cosine similarity between td-idf tokenized and vectorized tweet texts and their labels were compared based on the assumption that similar tweets are more likely to have a similar stance. Then, all our annotators were asked to discuss their annotation and the reasoning behind their decision, to reach a consensus on the inconsistent tweets. Thus, noisy labeling has been prevented. 

Some of the challenges we faced while annotating this data are: 
\begin{enumerate*}[label=\roman*)]
    \item Some of the tweets include irony/sarcasm; therefore, the true label might be hard to catch when the annotator is focused on the sentiment of the tweet only. 
    \item Since this debate is based on a health-related claim, the stances of a large amount of investigated tweets were ambiguous when the annotator only focused on the tweet text rather than a joint annotation of the tweet text and the shared URLs, if any. The source of this ambiguity is that many Twitter users shared URLs to the academic studies and news websites. 
    \item Each tweet can only contain up to 280 characters, which often poses a difficulty to fully understand the true meaning of the message, due to Twitter users being constrained to writing a short text. 
\end{enumerate*}
To overcome these challenges, we investigated the content of the URLs to understand the correct stances, only if the tweet text was not self-explanatory. 

\section{Annotation Guidelines}

We asked our annotators to label each of the tweets in our corpus using one of the three labels: \textit{"Against", "Favor", or "Neutral/None"} regarding the unproven claim of \textit{"Hydroxychloroquine and chloroquine are cures for the novel coronavirus."}. To gain a deep understanding of the stances in tweets, and due to the high subjectivity related to this classification method, the ambiguous tweets were being discussed in details among the annotation team. In the following sub-sections, the three classification labels and their corresponding tweet examples are explained in more details. 

\begin{enumerate}[label=\Roman*.]
    \item \textbf{AGAINST:} This stance label was being used for the annotation of tweets that imply an opposition to the claim, either directly or indirectly.

The stances of some of the tweets in this category are easily comprehensible as the tweet initiator expresses a direct opposition against the claim. For instance:
\textit{"I’m a physician. I would not take \#Hydroxychloroquine for \#COVID-19."}.

Some of the other tweets that were being identified as belonging to this category do not include personal opinions. Instead, these tweets might contain URLs to the academic studies or news articles in which hydroxychloroquine is demonstrated to be not effective against COVID-19, or simply contain the heading of the news article. It is assumed that the tweet initiator aims to share this information since he/she opposes the claim. An example of such a tweet is: 
\textit{"No evidence of clinical efficacy of hydroxychloroquine in patients hospitalized for COVID-19 infection with oxygen requirement: results of a study using routinely collected data to emulate a target trial | medRxiv"}.

Another example of expressing a counter attitude towards the aforementioned unproven claim, which was frequently observed in our data set, is via rationalizing against the claim. For instance, we observed that many Twitter users initiate contents that directly opposes the claim through expressing a reasoning, such as the side-effects of the drugs, or indirectly, via sharing the headings of news articles that imply the same concept:

\textit{"French Hospital Stops Hydroxychloroquine Treatment for COVID-19 Patients Over Major Cardiac Risk"},

\textit{"Mr. Trump himself has a small personal financial interest in Sanofi, the French drugmaker that makes Plaquenil, the brand-name version of hydroxychloroquine."}. 

Some tweets, on the other hand, include sarcasm/irony, which challenge the understanding of the true stance behind the textual content, even for human annotators. For instance:
\textit{"This is why you don't take your medical advice from a reality TV host. \#Hydroxychloroquine \#COVID19"}. 

    \item \textbf{FAVOR:} This stance label was being used for the annotation of tweets that imply a support opinion towards the claim, either directly or indirectly.

The stance label for some of the tweets in this category can be easily implied by the annotator, as the tweet initiator expresses a direct support in a straightforward language. For instance: 
\textit{"It is TIME to open up businesses and schools again! Corona numbers are inflated and you know it! And...there’s a cure!...hydroxychloroquine!"}

On the contrary to the above example, we observed that some of the tweet initiators convey their support of the claim in an indirect manner via questioning the conditions against what it seems to be an obvious solution to them. For instance:
\textit{Why did Fauci CHEER when hydroxychloroquine was used in 2013 for MERS, but is now skeptical for coronavirus?}

    \item \textbf{NEUTRAL/NONE:} This stance label is the last category in the annotation of our data set, and its label was being used for the tweets that are neither in favor, nor against the aforementioned claim. 
    
We observed that most of the tweets in this category fall into one of these groups: 
\begin{enumerate*}[label=\roman*)]
    \item tweets that are in the form of a question towards the truthiness of the claim, 
    \item tweets that convey a question with the aim of gaining more knowledge on the topic, 
    \item tweets that are written as a statement in a pure neutral tone, 
    \item tweets that only contain a neutral heading of a news article or an academic publication, and finally, 
    \item tweets that contain the query keywords; however, no clear relation between the claim under study and the tweet content were implied. 
\end{enumerate*}

An example of the tweets which convey a question as in group ii discussed above, is: 
\textit{"What is \#Hydroxychloroquine ??? How does it work a against \#Covid-19??"}

Finally, the tweets that are focused on the events related to the chloroquine/hydroxychloroquine drugs, rather than a direct relation to the claim under study is:
\textit{"India sends hydroxychloroquine to UAE for COVID-19 patients."}


\end{enumerate}

\section{Data Set Description}

\begin{figure}[!t]
\centering
\includegraphics[width = 1\linewidth]{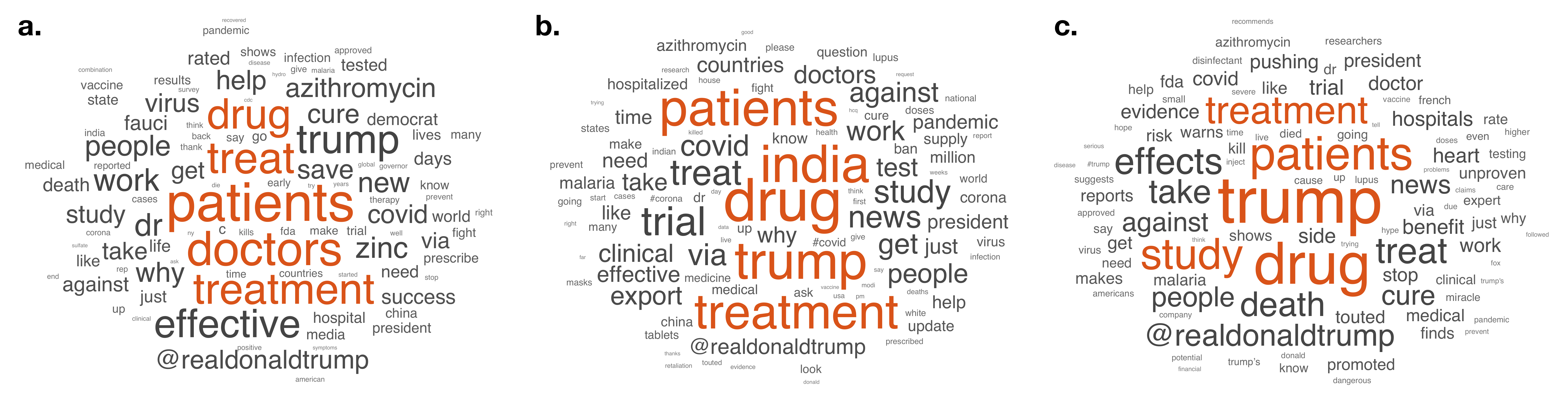}
\caption{This figure demonstrates the most frequently used keywords in each stance category: (a) favor, (b) neutral, (c) against.}
\label{fig:controversial_authors}
\end{figure}

The annotated COVID-CQ data set includes 14,374 original tweets (tweets/mentions/replies), which were generated by 11,552 unique users on Twitter. We excluded the retweets from our data set annotation. Table \ref{tab:stancedist} illustrates the size of our data set and the frequency of tweets for each class, in which the \textit{"Favor"} class with containing 6841 tweets is the largest class, followed by \textit{"Against"} class with 4685 tweets, and finally, the smaller class is \textit{"Neutral"} with 2848 tweets. 

\begin{table}[h!]
  \centering
  \caption{The frequency of the annotated tweets belonging to each stance class}
  \label{tab:stancedist}
  \begin{tabular}{cc}
    \toprule
   Stance&Number of Tweets\\\hline
    \midrule
    Neutral & 2848\\\hline
    Against & 4685\\\hline
    Favor & 6841\\\hline
    \textbf{Total} & \textbf{14374}\\\hline
  \bottomrule
\end{tabular}
\end{table}

To briefly provide information on the underlying topics in our corpus, we demonstrated the most frequently used words for each stance category in Figure \ref{fig:controversial_authors}. 
These word clouds are achieved after preprocessing the textual content of the data set, including the elimination of stopwords and the common domain words, such as chlorquine, hydroxychloroquine, covid19, and coronavirus. A detailed explanation of text preprocessing and cleaning is provided in section \ref{sec:assessment}. Despite the high similarity of the content in our corpus, and the intertwined topics for all the three clusters (i.e. the topics related to the available drugs, hospitalized patients, and the treatment methods), it is clearly observable that some of the keywords have been appeared in a specific stance class in a higher frequency. Since a considerable number of tweets in the \textit{Neutral/None} class are related to the import of the hydroxychloroquine drug from India to the US, the word ``India'' has appeared as one of the most frequent words in this class. For the \textit{Favor} stance class, we observed that plenty of tweets in this category are related to the effectiveness of hydroxychloroquine as combination with two other drugs, Zinc and Azithromycin. Thus, the word cloud for the \textit{Favor} class contains these keywords. Additionally, positive terms such as ``effective'', ``help'', ``save'', and ``success'' are also more observable in the textual content of this category. Finally, the \textit{Against} stance class has been observed to include words with a negative sentiment, including ``risk'', ``stop'', ``warn'', ``kill'', and ``death''.  

\begin{figure}[ht]
\centering
\includegraphics[width = 0.75\linewidth]{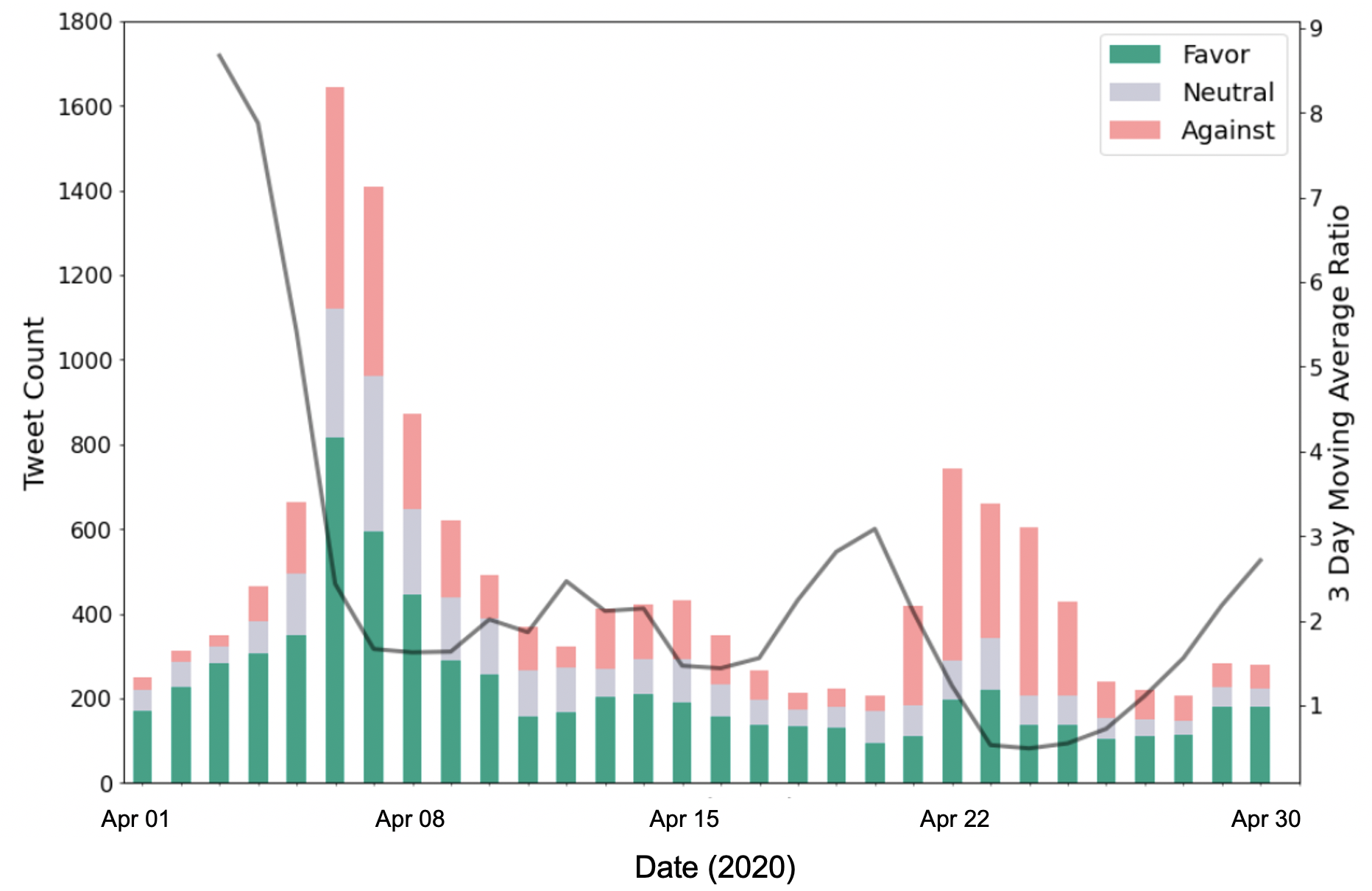}
\caption{The daily tweet counts in April 2020, classified into three categories: 'neutral', 'against', and 'favor'. The black line refers to the ratio of the number of 'favor' labeled tweets to the number of tweets with the 'against' stance, for a 3 day moving average.}
\label{fig:count}
\end{figure}

Figure \ref{fig:count} represents the daily counts of tweets that are labeled as \textit{"Favor"} (green), \textit{"Neutral"} (gray), and \textit{"Against"} (pink) in a one month time period. As the month of April is the beginning of the chloroquine/hydroxychloroquine debate, the fluctuation in the ratio between \textit{"Favor"} to \textit{"Against"} tweets are found as very drastic; the black line demonstrates the 3-day moving average of the ratio of \# of \textit{"Favor"} tweets to \# of \textit{"Against"} tweets. Therefore, this data set offers not only a challenging corpus for the stance detection task, but also presents a drastic dynamic for the researchers who focus on a better understanding of the information diffusion, polarization, and opinion changes over time, in their studies. 

The prepared data set is available to the public via our GitHub repository, accessible on \url{https://github.com/eceveco/COVID-CQ}. We adhere to Twitter’s terms and conditions by not providing the tweet JSON, but sharing the stance labels with the tweet IDs,  so  that  the tweets  can  be  rehydrated from the Twitter API.

\subsection{Major Events and Narratives}
Throughout the COVID-19 pandemic, many news articles and research publications are being discussed among social media users. 
Although many drugs made their path to mass clinical trials and researches across the world, the fierce debates surrounding the drugs hydroxychloroquine are more frequently observable among political figures, medical scientists, and social media users, while hydroxychloroquine is a less toxic metobolite of chloroquine and has been identified to have less side effects \cite{stokkermans2019chloroquine}. 
As the COVID-CQ data set is focused on a month-length Twitter activities regarding the COVID-19 pandemic, and particularly, on the chloroquine/Hydroxychloroqune conversations, this data contains textual content in relation to many major events that occurred since the beginning of the pandemic up to the end of April 2020. As discussed in \cite{toktam2020}, narrative summaries can be constructed from an ordered chain of individual events with causality relationships amongst events, appeared within a specific topic. According to this definition, below we briefly narrate the major events that are being discussed in the Twitter conversations in our data set. However, this narration does not reflect authors' personal opinions towards any of the reviewed events. 

Amongst the narratives that are discussed in this data set while being occurred before April is the news related to the results of a study published on March 20 regarding hydroxychloroquine to treat COVID-19 patients, which found that treating patients with a combination of hydroxychloroquine and azithromycin results in a more efficient virus elimination  \cite{gautret2020hydroxychloroquine}. This combination of the drugs has been referred to as ``game changer'' and ``beginning of the end of the pandemic'' by many Twitter users. However, plenty of studies later failed to replicate these results \cite{geleris2020observational}. 

Examples of other events that attracted significant attention from Twitter users are the purchase of hydroxychloroquine sulfate tablets by the Department of Veterans Affairs and the Bureau of Prisons in March 2020. We observed in our data that the tweet initiators who propagated news on these events were referring to many news articles regarding this narrative, including the articles published by New York Post\footnote{nypost.com/2020/04/07/federal-agencies-purchase-large-supply-of-hydroxychloroquine/} and The Daily Beast\footnote{thedailybeast.com/the-bureau-of-prisons-just-bought-a-ton-of-hydroxychloroquine-trumps-covid-19-miracle-drug?ref=scroll}. Further discussions in regards to the purchase and storage of these drugs in our data set are related to the US government announcement of stockpiling the drug hydroxychloroquine in late March, the event that according to Bloomberg\footnote{bloomberg.com/news/articles/2020-03-20/hospitals-stockpile-malaria-drug-trump-says-could-treat-covid-19} was later followed by many hospitals in the United States. Although a substantial number of users on social media only shared the URLs to news article and/or included the headings on these topics, many individuals argued that the mass storage of the unproven drugs by the US administration and hospitals might result in the deprivation of patients in true needs to access the drugs, such as the Lupus patients who receive antimalarial drugs to ease their symptoms \cite{ponticelli2017hydroxychloroquine}. A Bloomberg article\footnote{usatoday.com/story/news/health/2020/04/18/hydroxychloroquine-coronavirus-creates-shortage-lupus-drug/5129896002/} refers to this issue as the shortage of the drug for Lupus patients, as a consequence of its high demand for COVID-19.  

Another important event in March 2020 that affected the opinions on online social media was the Emergency Use Authorization (EUA)\footnote{fda.gov/media/136534/download} for oral formulations of hydroxychloroquine sulfate and chloroquine phosphate by the Food and Drug Administration (FDA), granted on March 28. We observed that this event attracted significant attention on Twitter during April, and that it encouraged the appearance of a positive attitude towards the drugs. 

On March 25, India that is one the largest manufacturers of the drug hydroxychloroquine, announced that the directorate-general of foreign trade has prohibited the export of this drug to any other countries amid coronavirus outbreak \footnote{statnews.com/pharmalot/2020/03/25/india-trump-hydroxychloroquine-coronavirus-covid19/}. This decision was partially lifted on April 5 in response to a call from president Trump to Indian Prime Minister Narendra Modi, according to BBC\footnote{bbc.com/news/world-asia-india-52196730}. Many Twitter users propagated the news on the lift of export ban for this drug by using \#retaliation. 

In early April, the results from various clinical trials related to at least 15 different treatments were announced by many news agencies. Among the news articles, an international poll from 2,171 physicians started to propagate in Twitter conversations, in which 37\% of doctors rated the drug hydroxychloroquine as the most effective therapy to combat the virus. This report was also reflected by the New York Post\footnote{nypost.com/2020/04/02/hydroxychloroquine-most-effective-coronavirus-treatment-poll/} on April 2.  

On April 7, another rumor emerged, but this time regarding the immunity of patients with rheumatology illnesses against the novel coronavirus, as a result of taking the drug hydroxychloroquine. 
The Twitter users started to propagate this rumor after the rheumatologist, Dr. Daniel Wallance, mentioned in an interview with Dr. Oz on April 7 that out of 800 patients who are taking the drug, none have been reported to contract the virus \footnote{youtube.com/watch?v=kd7Jec3pZBk}. However, the controversial reports appeared from late April, in some the patients with Lupus were being considered as to be at higher risks of infection and development of severe symptoms for coronavirus\footnote{medicalnewstoday.com/articles/lupus-and-covid-19}.

The support of hydroxychloroquine during the white house press briefings on the coronavirus pandemic influenced the opinions on Twitter negatively. Consequently, social media users started to relate political and/or financial benefits to the support of the drug, which was also being discussed in many news articles, including what Washington Post calls ``the real reason behind hydroxychloroquine obsession''\footnote{washingtonpost.com/opinions/2020/04/07/real-reason-trump-is-obsessed-with-hydroxychloroquine/}, published on April 7.

Despite the positive news regarding the efficiency of the drugs until the middle of April 2020, the results published in a study of hundreds of patients at US Veterans Health Administration medical centers suggested that patients taking hydroxychloroquine are no less likely to get infected by the virus, instead, the death rates in these peatiest have been observed to be higher \cite{magagnoli2020outcomes}. These results appeared in many news articles, including a CNN article\footnote{cnn.com/2020/04/21/health/hydroxychloroquine-veterans-study/index.html} published on April 21. 

With respect to the the negative side effects of hydroxychloroquine that were reported in many research articles, the FDA issued a warning about the use of this drug for COVID-19 patients on April 24, according to Time\footnote{time.com/5827085/fda-warning-hydroxychloroquine/}. The severe negative effects reported for the use of this drug include abnormal heart rhythms that might threaten patients' lives. 

The FDA warning related to the use of antimalarial drugs to treat COVID-19 patients and the publication of academic researches which reported the inefficacy of these drugs against the novel coronavirus caused the United States to be left with massive supplies of hydroxychloroquine, the concern that was being reflected in many news websites at the end of April, including the article published in USA Today\footnote{usatoday.com/story/news/politics/2020/04/27/coronavirus-states-stockpile-hydroxychloroquine-drug-trump-touted/3031660001/} in April 27. Finally, in June 2020 the FDA withdrawn the granted emergency use authorization (EUA) for the drug hydroxychloroquine\footnote{fda.gov/media/136534/download}. However, the EUA withdrawal event does not appear in our data set.

\section{Annotation Assessment}
\label{sec:assessment}
After the preparation of the data set according to the annotation guidelines, we conducted extensive analysis on the data to ensure the quality of the annotation, including the consistency of the stance labels in our corpus. In this regard, we investigated the semantic similarity of the tweets via computing the pairwise cosine similarity on the vector representation of the tweets. The Universal Sentence Encoder\footnote{\url{tensorflow.org/hub/tutorials/semantic_similarity_with_tf_hub_universal_encoder}} from the TensorFlow library was used to achieve the semantic vector representations of the tweets. The sentence level embeddings provide high level sentence semantic relationship, which enables the comparison of the similarity of tweet contents against each other to assure labeling consistency. After the calculation of pairwise cosine similarity for the tweets in our corpus, we reevaluated the labels of the tweets with $\geq0.9$  cosine similarity, where this threshold was being identified by human judgement via comparing the tweet similarity results. The reevaluation procedure of the identified highly similar tweet contents include manual investigation of these tweets by the annotators, such that the highly similar tweets that convey the same stance towards the claim are categorized into the same stance class.   

\subsection{Stance Classification}
To demonstrate the potential of evaluating many stance classification models using this data, and to evaluate the quality of our data set, we conducted extensive analysis using six different classification methods. The implemented models for this purpose include Multilayer Perceptron (MLP), Logistic Regression (LR), Support Vector Machine (SVM), Multinomial Naive Bayes (MNB), Stochastic Gradient Descent (SGD), Gradient Boosting (GB), and finally, Convolutional Neural Network (CNN). Furthermore, all the classification methods are compared for the computed vector representations of word unigrams and bigrams using the term frequency-inverse document frequency (tf-idf). 

The implemented Multilayer Perceptron contained 2 dense layers, the rectified linear unit (ReLU) as the activation function, and the cross-entropy loss as the loss function. For the Logistic Regression classifier, the Limited-memory Broyden Fletcher Goldfarb Shanno (lbfgs) was used as the solver. The Support Vector Machine model was implemented with the linear kernel. For the Stochastic Gradient Descent model, the perceptron loss was used. In the implementation of the Gradient Boosting model, the deviance loss was used for model optimization. Finally, the Convolutional Neural Network (CNN) was implemented in two different ways to receive the inputs as vectors computed by one-hot encoding and GloVe word embedding, both with 5 convolutional layers with kernel size of 3 and stride size of 2, and with Exponential Linear Unit (ELU) activation function. The training stop criteria was to 
reach to a maximum number of 1000 iterations in training for all the classifiers. 

\subsection{Data Preprocessing}
To prepare the input to the classifiers, we first filtered the tweets that were identified by Twitter to be in a different language than in English. After excluding the non-English tweets, we removed any punctuation marks and non-Ascii characters, and replaced the integers with their textual representation. Further preprocessing of the data include mapping all the input text to lowercase format, followed by word stemming, lemmatization, and the removal of the stopwords. Additionally, the URLs, hashtag signs ($\#$\textit{xxx}), and emoticons were removed to achieve a higher textual quality. It should be noted that none of these classification methods have been trained or tested on the content of the shared URLs as part of the input data. After this step, the term frequency-inverse document frequency (tf-idf) was used to generate the vector representation of the input tweets. Using tf-idf, we generated a vector space with weighting scheme based on the frequency of unigrams and bigrams in a tweet relative to the total number of their frequencies in the entire data set. Thus, tf-idf captures the most distinct words while ignoring the semantic or syntactic attributes. After this step, the vectorized tweets were used as the input to all the classifiers. For the training and testing of all the models, we used 80\% of the tweets in training, and the remaining of the tweets to evaluate the models. 

\subsection{Results}

\begin{table}
\centering
\caption{The comparison of the results for 6 classifiers using tf-idf vectorization}
\label{tab:classifiers}
\begin{tabular}{l|l|l|}
\cline{2-3}& \multicolumn{1}{c|}{Unigram} & \multicolumn{1}{c|}{Bigram} \\ \cline{1-3} 
\multicolumn{1}{|l|}{Stochastic Gradient Descent (SGD)}                & 0.7429     & 0.7439     \\ \hline
\multicolumn{1}{|l|}{Support Vector Machine (SVM)}                      & 0.7651     & 0.7651     \\ \hline
\multicolumn{1}{|l|}{Multilayer Perceptron (MLP)}                       & 0.7453     & 0.7457     \\ \hline
\multicolumn{1}{|l|}{Logistic Regression (LR)}
        & \textbf{0.7683}     & \textbf{0.7683}     \\ \hline
\multicolumn{1}{|l|}{Multinomial Naive Bayes (MNB)}   
        & 0.7182     & 0.7182     \\ \hline
\multicolumn{1}{|l|}{Gradient Boosting Classifier (GB)}
        & 0.6764     & 0.6768     \\ \hline
\end{tabular}
\end{table}

The comparison of the results for 6 classification models using our stance data set is provided in table \ref{tab:classifiers}. Among the investigated classification methods, the Logistic Regression model achieved the best accuracy of 0.76 for both accuracies when the feature vectors for the tweets were computed using unigrams and bigrams for tf-idf. The next best performance was achieved by the SVM with very close performance to LR. The gradient boosting model achieved the lowest performance for both unigram and bigram tf-idf vectorized inputs. Surprisingly, using the bigrams to generate the tf-idf feature vectors did not affect the accuracy, observed for all models. However, despite the use of general purpose classification methods than state-of-the-art stance detection models, and although the contents from the URLs were not used for the evaluation of these classifiers, all of the models were able to classify the tweets with an acceptable performance. For further analysis, we implemented a Convolutional Neural Network (CNN) classifier for stance classification with one-hot encoding and GloVe vectorization of the words in tweets. This classifier achieved the accuracy of 0.73 for both vectorization methods. Additionally, the stance classification using the MLP model was also repeated for the one-hot encoded tweets. The achieved classification accuracy for this classifier was 0.75, which is slightly improved comparing with the MLP model using the tf-idf feature vectors.

\section{Discussion}
In this work, we introduced a large data set of Twitter stances towards the unproven claim of ``chloroquine and hydroxychloroquine are cure for the new coronavirus''. Our data set, COVID-CQ, contains stance labels for more than 14 thousand original tweets, after discarding the retweets. COVID-CQ defers from the existing corpus as the true underlying stances in Twitter conversations have been identified via a joint annotation of tweets' text and the shared URLs when the tweets were not self-explanatory. Accordingly, our data set challenges the prediction models for the tasks of stance detection, to incorporate further information besides the text of the tweets. We have made the annotated corpus available to the public through our GitHub repository, in which the Tweet ids and the stance labels are provided to the research community and the given information can be used for Tweet rehydration via the Twitter API. To the best of our knowledge, COVID-CQ is the first data set regarding the stances towards the COVID-19 pandemic, besides being the largest human annotated stance data set for social media on stances towards a claim.

\section*{Acknowledgments}
The authors gratefully thank Mina Sonbol and Nicholas Wiesenthal who assisted in the annotation and analysis of the data set.

\bibliographystyle{unsrt}  
\bibliography{references}
\end{document}